\documentclass[reprint,
 amsmath,amssymb,nofootinbib,aps,
]{revtex4-1}
\usepackage{xcolor,color}
\usepackage[utf8]{inputenc}
\usepackage{float}
\usepackage[T1]{fontenc}
\usepackage{latexsym,amssymb,amsmath,amsfonts}
\usepackage{subfigure}
\usepackage{graphicx}
\usepackage{mathrsfs}
\usepackage{float}
\usepackage{enumerate}

\begin{document}

\title{Extended error threshold mechanism in {\it quasispecies} theory via population dynamics}

\author{Hermano Velten}%
 \altaffiliation[]{hermano.velten@ufop.edu.br (corresponding author)}
 \author{Carlos Felipe Pinheiro}
 \author{Alcides Castro e Silva}
%\email{Second.Author@ins
 %\email{Second.Author@ins
\affiliation{%
$^{1}$Departamento de F\'isica, Universidade Federal de Ouro Preto (UFOP), Campus Universit\'ario Morro do Cruzeiro, 35.402-136, Ouro Preto, Brazil}
\date{\today}% It is always \today, today,

\begin{abstract}

We investigate Eigen's model for the evolution of the genetic code of microorganisms using a novel method based on population dynamics analysis. This model, for a given number of offspring, determines long-term survival as a function of the "genetic" information length and copy error probability. There exists a maximum threshold for the quantity of information that can be consistently preserved through the process of evolution within a population of perfectly replicating sequences, meaning no errors are allowed. With our formula, we expand upon the traditional error threshold formula of Eigen's theory and introduce a new expression for general cases where the self-reproduction process allows up to any integer number of copying errors per digit per replication step.

\end{abstract}

\maketitle

\section{Dynamics of self-replicating systems: the Eigen model}

\subsection{Overview}

Self-replication is a necessary condition for life and a building block of {\it de novo} synthesis. The latter refers to the creation or assembly of complex molecules or structures from simple building blocks or basic components, synthesizing them without relying on pre-existing materials. During the process of {\it de novo} synthesis, two essential aspects come into play. First, the organism must have the capability to produce precise and faithful copies of itself to achieve self-replication. This accuracy ensures that subsequent generations maintain the genetic information and functional characteristics of the parent organism. The ability to faithfully replicate genetic material is crucial for the perpetuation and continuity of life. Second, the replication process is not entirely error-free. Mutations or errors can occur during DNA replication, giving rise to genetic variations. These variations, in turn, provide the raw material for Darwinian evolution and the emergence of diverse traits within a population. Without these genetic variations, natural selection would have limited material to act upon, hindering the adaptation and survival of species over time.

To ensure the preservation of accurate genetic information while allowing for genetic variations, error correction mechanisms play a vital role. These mechanisms act as quality control systems, identifying and repairing errors that arise during replication. By minimizing the occurrence of errors or mutations, these correction mechanisms maintain the fidelity of genetic information, allowing the stability and continuity of species.

Thus, in the {\it de novo} synthesis process, the organism's ability to generate accurate copies of itself supports self-replication, while the presence of error correction mechanisms provides the necessary framework for Darwinian evolution and the ongoing diversification and adaptation of life forms.

The first mechanism to link  biological information to physico-chemical reactions via a kinetic theory for replication processes was ellaborated by Manfred Eigen \cite{Eigen1971}. In this scheme, the term "error threshold" refers to the upper limit or critical point in the length of an information sequence that a self-replicating system can support and it is defined in relation to the error or mutation rate during the reproduction process.

In the Eigen model, self-replication is accompanied by a certain error rate, where errors or mutations can occur during the replication of genetic information. These errors can lead to imperfect copies or variations in the offspring's genetic sequences. The concept of error threshold focuses on the balance between accurate replication and accumulation of errors (Refs. \cite{Biebricher,Summers, tarazona1992error} also discuss this concept).

When the error rate exceeds a certain threshold, the errors accumulate rapidly, resulting in a deteriorating fidelity of replication. This leads to a point where the information becomes increasingly distorted and the offspring's viability declines. In such cases, the self-replicating system may fail to maintain accurate genetic information, leading to the elimination of the offspring and the eventual collapse of the population.

The error threshold can be seen as a critical value where the self-replicating system transitions from a state of sustainable replication to one of error catastrophe. Below the threshold, the system can sustain long-living populations, maintaining genetic stability and enabling the persistence of favorable traits. However, beyond the threshold, the system becomes increasingly error-prone, leading to a decline in replication fidelity and the potential loss of functional genetic information. Understanding the error threshold is important once it provides insight into the conditions necessary for the long-term survival of self-replicating systems. It highlights the delicate balance between accurate replication and the occurrence of errors, ultimately shaping the dynamics of evolutionary processes in populations.

Let us define the chain length of the information sequence $L$ supported by a self-replicating system with a single-digit error rate $p$. The single realization of an error, or mutation, during the reproduction process in any information unit is understood as an imperfect copy leading to elimination of the offspring. For self-replication process giving rise to $\sigma$ offsprings the so-called ``Eigen'' condition for successful long-living population is \cite{Eigen1971,EigenSchuster}

\begin{equation}
L < \frac{\mathrm{ln} \,\sigma}{p}.
\label{1}
\end{equation}

The expression above (\ref{1}) originally appeared in \cite{Eigen1971,EigenSchuster} as a direct consequence of information theory. Eigen's model posits that the successful reproduction of a quasispecies hinges on ensuring the stable conservation of information during the process of selection. The combination of selection dynamics with the informational aspect, as characterized by Shannon's entropy, results in a threshold relationship governing the maximum information stored in the evolutionary dynamics of a quasispecies \cite{Bull2005,domingo2012viral,adami2004information}.

Since the investigation of mathematical aspects of the quasispecies theory is a firm line of research (interesting mathematical results can be found in Refs. \cite{fontanari,Nilsson,Park,Saakian} and references therein and also reviews in Refs. \cite{Domingo,Baake,Bratus2019}), we follow in this work this reasoning by extending the the Eigen's mechanism to the population dynamics language. Populational dynamics is definitely a more familiar tool within evolutionary biology studies than the intricate mathematical aspects present in the original Eigen's proposal. This is also supported by arguments stating that there is complete consistence between quasispecies theory and population genetics \cite{Wilke}.

\subsection{Eigen relation and population dynamics}

In order to illustrate the Eigen relation (\ref{1}) let us take the example of a simple organism with information encoded in a string of length 
$L=2$. Then, there are $2$ possible information units in which information can be stored. During the copying process, in case the information from each location is reproduced to the next generation, this organism generates, let us say, for instance, $\sigma=2$ offsprings. Then, after reproduction, the total population is now $N=2$ since the original organism no longer exists. Population grows obeying a geometrical progression as long as there is no copy error at any of the $L=2$ locations. For every new generation, the error probability $p$ acts in each of the $L=2$ digits. Let us admit now that the information contained within each of the $L$ information units is binary (one can wonder this binary structure as activating/deactivating some biochemical process). The possibilities for an organism with chain $(0,0)$ are i) it does not suffer a mutation $(0,0)$, ii) it suffers a single mutation, becoming $(1,0)$ or $(0,1)$, or iii) it suffers a double mutation $(1,1)$. If the probability of error for each information unit is $p$ then the probability that an error does not occur is straightforwardly $q=1-p$ and the chance the organism is not affected by the error error is $qq=q^2$. The chance of only one error is $pq+qp=2pq$ while two error probability reads $pp=p^2$. It is worth noting that the total probability of being or not being affected by a copying error is $p^2+2pq+q^2=(p+q)^2\equiv 100 \%$. For our simple example, if $p=0.1$ then $q=0.9$ and the probability of not suffering a mutation is $q^2=0.9^2=0.81$, while the probability of a single mutation is $2pq=0.18$ and two mutations $p^2=0.01$. The total chance of having a copy error is $p^2+2pq=0.19$. As another example, for an organism with $L=5$ the equality $(p+q)^5=p^5+5p^4q+10p^3q^2+10p^2q^3+5pq^4+q^5=1$ represents the sum of all probabilities evolved in the copy process. Except from the term $q^5$, representing the case of no copy error, all other terms represent one of the copy error possibilities in different $L$ positions \footnote{All such probabilities can be calculated from Newton binom $(p+q)^L=\sum_{M=0}^L \left( \begin{array}{c} L \\ M \end{array} \right) p^{L-M} q^M$, and $\binom{L}{M} = \frac{L!}{M!(L-M)!}$.}. In this case the error probability in one information unit is $5pq^4=0.33$. When comparing the cases for $L=2$ and $L=5$ the probability of error in a single information unit changed from $18\%$ to $33\%$. Taking into account all possible mutations that can take place for the case $L=2$ one has $p^2+2pq=0.19$ while for the case $L=5$ one has $p^5+5p^4q+10p^3q^2+10p^2q^3+5pq^4=0.41$. This exemplifies that more complex organisms are more sensitive to copy errors, forcing them to design evolutionary mechanisms that decrease the $p$ value.

\begin{figure}[ht]
\centering
\includegraphics[scale=0.41]{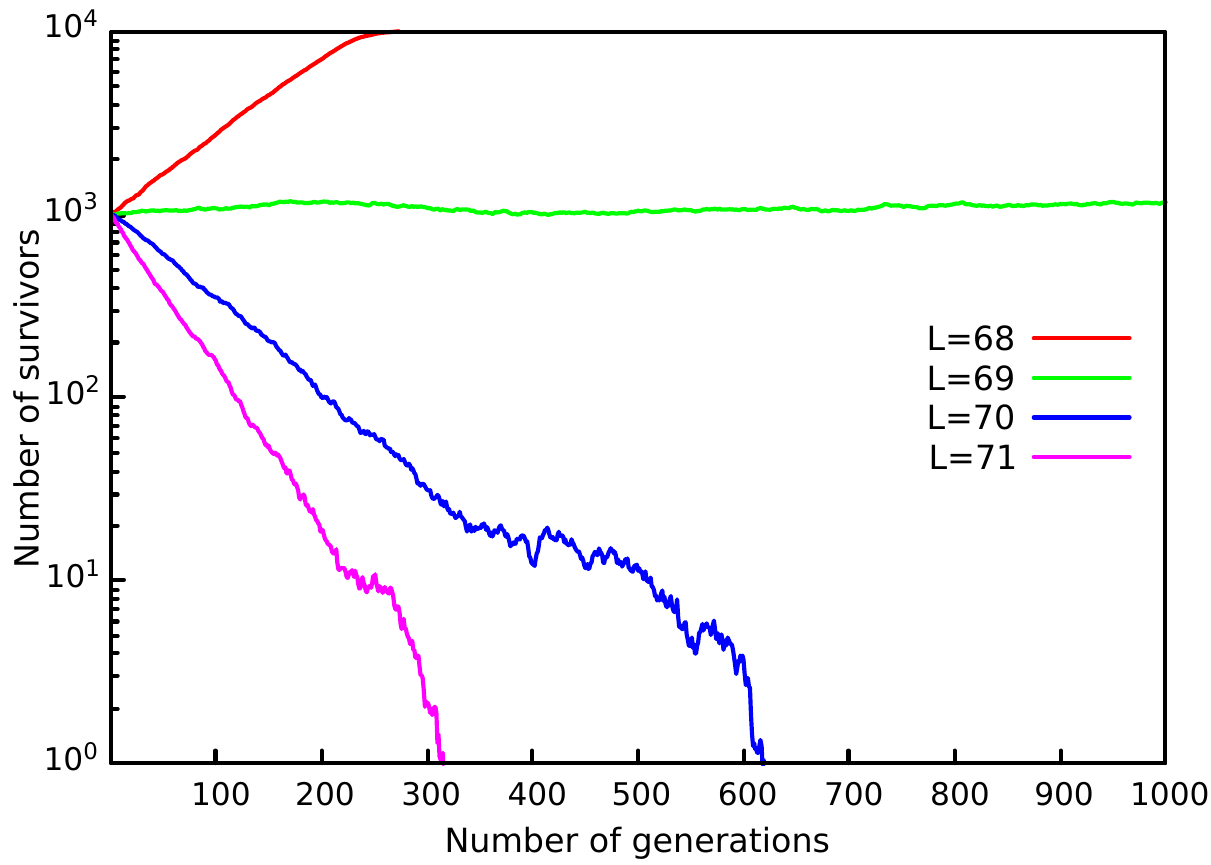}
\caption{Visualization of criteria (\ref{1}) for long lived replication steps. This figure shows the evolution of $4$ distinct populations with information lengths $L=68, L=69, L=70$ and $L=71$. Mutation rate has been fixed to $p=0.01$, number of offsprings $\sigma=2$ and initial population consisting in $1000$ organisms. According to (\ref{1}), since $ln (2) = 0.693$ and $p=0.01$ the threshold is achieved for $69<L<70$. Simulation based on the discussion proposed in Ref. \cite{Brandenburg:2008uh}.}
\label{figure1}
\end{figure}

In Fig. \ref{figure1} we show the simulation for $3$ different populations with different lengths of information codes, namely $L=68$, $L=69$, $L=70$ and $L=71$ with initial population size $N(t=0)=1000$. In this simulation each reproductive cycle gives rise to $\sigma=2$ offsprings. Larger populations accumulate more errors fading away earlier. According to Eq. (\ref{1}) the threshold occurs when $pL>0.693$. The green line ($L=69$) indeed decays for very large generation number. 
 
For a generic organism with chain length $L$ one can calculate the probability $P_M$ as a function of the minimum number $M$ of errors admitted having a mutation probability $p$ in each information unit every replication process.

For example, if the organism is eliminated with a single error in any information unit $P_1=1-(1-p)^L$. In case it supports one error but dies with two or more errors $P_2=1-(1-p)^L-Lp(1-p)^{L-1}$.

It is easy to predict how the relation between $P_1$ and $p$ behaves. Although $P_1$ increases with $p$ in large chains, $P_1 \rightarrow 1$ fastly according to the fact that more complex organisms are more sensitive to error rates. Complexity has been possible via the development mechanisms that minimize consequences of high mutation rates allowing the development of large coding chains. Then, simple organisms have maintained their populations long enough to assure the appearance of such evolutionary strategies.

Let us now try a different approach to obtain the Eigen relation (\ref{1}). We start from typical populational dynamics arguments i.e., elaborating a recurrence relation for the population variation $\Delta N$ from the moment $t$ until a subsequent instant $t+\Delta t$. Let $N(t)$ represent the population at a given time $t$, and $\Delta t$ be the time until the subsequent generation emerges. If any single individual gives rise to $\sigma$ elements, then one has $\sigma N(t)$ new organisms. Having only new elements the variation between time $t$ and $t+\Delta t$ would be $\Delta N = \sigma N(t)$. However, the original organism is eliminated in this process leading to new $\sigma$ elements ($N(t)$ eliminations). Also, one has to consider the $P_1 \sigma N(t)$ deaths by copying error. The index $M$ and $P_M$ indicates that the organism resists to only $M$ mutations every cycle. Therefore the variation rate in the number of elements becomes:
\begin{equation}\label{eq:deltaT}
	\Delta N = \sigma N(t) - N(t) - P_1 \sigma N(t).
\end{equation}
Since $\Delta N = N(t+\Delta t) - N(t)$ and the probability of occurrence of at least one copy error is $P_1 = 1 - (1-p)^L$ one finally finds
\begin{equation}\label{eq:iterativo}
	N(t+\Delta t) = (1-p)^L \sigma N(t).
\end{equation}

Eq.~(\ref{eq:iterativo}) is actually the interactive map of our dynamics. Starting from a population $N(0)$ and applying this relation recursively one obtains the total population at any point in the future. In Eq. (\ref{eq:iterativo}) is also possible to assess whether the population grows depending on $L$ and $p$. The condition for growing population is $(1-p)^L \sigma > 1$,  or {\it mutatis mutandis}, 
\begin{equation}\label{eq:pre-eigen}
L \ln (1-p) > - \ln \sigma.
\end{equation}
Multiplying both sides of Eq.(\ref{eq:pre-eigen}) by $-1$ and using that $p\ll 1$, $\ln (1-p) \approx -p$
one arrives at the conclusion (\ref{1}) i.e., $pL < \ln \sigma$.

\subsection{Population dynamics consistency}

Although the Eigen's relation was originally obtained via information theory, it is completely consistent with population dynamics \cite{royama2012analytical}. 
In Eq.~(\ref{eq:iterativo}) it is easy to find a condition when the population shows a stationary dynamics, that is, the number of grows matches the number of deaths due mutation. In this case the population reachs a fixed value given by $N(t+\Delta t)=N(t)=N^{\star}$, where $N^{\star}$ is called a fixed point. Introducing that condition in Eq. \eqref{eq:iterativo}
we have:
\begin{equation}
    [(1-p)^L\sigma-1]N^{\star}=0.
\end{equation}
This relation has two solutions, a trivial one where $N^{\star}_1=0$, or extinction, and $(1-p)^L\sigma-1=0$, leading to Eigen's relation.

Let us ignore for one moment the mutation parameter, doing $p=0$ and introduce a carrying capacity of the environment $K$. This is an important concept used in ecology and refers to the maximum population size a specific environment can sustain indefinitely \cite{del2004carrying}.  Eq.~(\ref{eq:iterativo}) can be rewritten as
\begin{equation}
    N(t+\Delta t)=\sigma N(t)-P_K\sigma N(t),
\label{N(t+Dt)}
\end{equation} 
where $P_K$ is a probability of death due a environmental overload, that means, $P_K=N(t)/K$.
Our equation \eqref{N(t+Dt)} turns into 
\begin{equation}
N(t+\Delta t)=\sigma N(t)-\frac{\sigma}{K} N^2(t)=\sigma N(t)[1-N(t)/K], 
\end{equation}
as known as the Logistic Map.
Analysing fixed points in this new equation we have
\begin{equation}
  \left[(1-\sigma)+\frac{\sigma N^{\star}}{K}\right]N^{\star}=0 . 
\end{equation}
Once again we have here a trivial fixed point $N^{\star}_1=0$, but now there is a second well defined fixed point when $[(1-\sigma)+\sigma N^{\star}_2/K]=0$, or $N^{\star}_2=K(1-\sigma^{-1})$.
What happens adding mutation $p$ and carrying capacity? The new iterative map becomes
\begin{equation}
    N(t+\Delta t)=(1-p)^L\sigma N(t)-P_K \sigma N(t).
\end{equation}
With the above definition for $P_K$ it reads then
\begin{equation}
    N(t+\Delta t)=\sigma N(t)\left[(1-p)^L-\frac{N(t)}{K}\right].
\end{equation}

This last map has again two fixed points, a trivial $N^{\star}_1=0$ and a non trivial $N^{\star}_2=K[(1-p)^L-1/\sigma]$. It is worth noting that $N^{\star}_2$ must be positive, so $(1-p)^L> 1/\sigma$ leads again to Eigen's relation.

\section{Error accumulation}

For complex organisms with $L\sim 10^8$ the Eigen condition is satisfied if the copy error of order $p\lesssim 10^{-8}$. Simpler chains e.g.,  $L\sim 10^{4}$, can keep the population alive with larger error probabilities. Then, modern organisms have evolved by improving the reproductive process by decreasing copy error rates. 

Now, we try to model such adaptation methods to the Eigen condition. Let us extend the previous relations to the situation where the sequence keeps growing even having $M-1$ copy errors per digit where $M$ is an integer number. In other words, lethality comes when errors start to accumulate (mutation-accumulation is the key idea behind the Evolutionary Theory of Aging or Senescence, proposed by Peter Medawar \cite{charlesworth2000fisher}).

We start rewriting it such that
\begin{equation}\label{eq:dinamica-P1}
	N(t+\Delta t) = (1-P_M) \sigma N(t).
\end{equation}
Then, what is the difference when the organism acquires the ability to support more than a single error? 

Let us start by supposing that the organism dies when accumulates $2$ errors in the self-replication step, but it can live with one. Hence, such probability becomes
\begin{equation}
    P_2 = 1 - (1-p)^L - Lp(1-p)^{L-1}.
\end{equation} 

In this scenario, the probability $P_1$ has to be replaced by $P_2$ in Eq.~(\ref{eq:dinamica-P1}).

Eq. (\ref{1}) gives the condition for long-living population when even one error in replication causes death, in other words, there is no error accumulation whatsoever. If an organism evolves in order to support one error and dying with two, the Eigen relation must be rewritten.
This can be done starting with Eq. (\ref{eq:dinamica-P1}), but changing $P_1$ by $P_2$. That way we have
\begin{equation}
    N(t+\Delta t)=[(1-p)^L+Lp(1-p)^{L-1}]\sigma N(t),
\end{equation} 
and the new survival condition is
\begin{equation}
    [(1-p)^L+Lp(1-p)^{L-1}]\sigma>1.
\end{equation}

Expanding the left hand side in Taylor series until $O(2)$ and rearranging the expression one finds 

\begin{equation}
    p^2L(L-1) < 2~\ln\sigma.
\end{equation}

Keeping the same reasoning as we did above, by extending this analysis to any integer $M$, given in terms of the binomial notation $P(E) = {n \choose k} p^k (1-p)^{ n-k}$, we finally find the general expression for the error threshold formula:
\begin{equation}\label{maineq}
    {L \choose M} ~p^M <~\ln\sigma,
\end{equation}
valid for any $M$ integer. Equation \eqref{maineq} stands as a general expression of \eqref{1}. Similarly to the original relation, it defines conditions for survival of a quasispecies. However, in this generalization it is introduced the parameter $M$ that is a threshold of how many ($M-1$) mutations that organism can accumulate. As an example, a quasispecies with $M=3$ can support $2$ mutations, but it dies with the $3rd$ one. For $M=1$, equation \eqref{maineq} turns into \eqref{1}.

That way, we are introducing the extended threshold formula admitting the existence of non lethal mutants (with $M-1$ mutations), what is essential if one wants to define {\it fitness} criteria in darwinian evolution scenarios.
It is expected that organisms with $M>1$ can maintain viable populations for a higher values of $p$ when compared to a population with same $L$ but under Eigen's condition $(M=1)$. 

In order to quantify how this probability changes with $M$, let us write down the rate $\mathcal{P}_M\equiv P_M/P_1$ for a fixed $L$ population:
\begin{equation}\label{pp}
\mathcal{P}_M = \frac{\left[\dfrac{\ln(\sigma)}{{L \choose M}}\right]^{1/M}}{\dfrac{\ln(\sigma)}{L}} = \frac{L}{{L \choose M}^{1/M}} \times \frac{1}{\ln(\sigma)^{1-1/M}}.
\end{equation}

For the expect limiting case when $L>>M$, we have:
\begin{equation}\label{pp2}
    \mathcal{P}_M \approx \sqrt[M]{\frac{M!}{\ln(\sigma)^{M-1}}}.
\end{equation}
Equation \eqref{pp2} shows how a single-digit mutation rate $p$ defined in Eigen's scenario can increase and still maintain viable populations as a function of $M$. As example, for $M=2$, we have $\mathcal{P}_2 = 1.69$, for $M=5$, $\mathcal{P}_5 = 3.49$. When $M=1$ the result is $1$ as expected. It is remarkable that equation \eqref{pp2} does not depend on $L$, showing that $\mathcal{P}_M$ saturates for large $L$. That suggests there is a limit where getting the sequence $L$ larger brings none advantage dealing with mutation rates despite the energy cost in growing the machinery evolved on the process.

\section{Final Discussion}

After the publication of Manfred Eigen’s seminal paper \cite{Eigen1971}, there has been extensive examination of the mathematical peculiarities inherent in the quasispecies model. The present work also addresses another interesting mathematical aspect of this theory. We have presented a population dynamics approach to the quasispecies model. By applying population dynamics techniques to the error threshold formula, we provide a simpler expression for cases where the self-reproduction process supports any integer number $M-1$ of copying errors per digit per replication step. This is our main result, presented in \eqref{maineq}.

Let us apply this result to well-known organisms. Although the most ancestral replicable molecule is unknown, several lines of evidence suggest that RNA preceded DNA. RNA is a single-stranded chain, whereas DNA is double-stranded. There is a hypothesis that in the early stages of life, RNA was used as the hereditary molecule, and DNA came to be used later \cite{joyce2002antiquity, saito2022rna,neveu2013strong, orgel2003some}. Current RNA viruses, such as HIV, have a size on the order of $10^4$ nucleotides, whereas DNA-based organisms have much larger chains, such as $10^6$ for bacteria, $10^8$ for worms like {\it Caenorhabditis} elegans, and $10^9$ for humans \cite{ridley2013evolution}. In case quasispecies can be used to RNA replication (see \cite{Moya} for contrary claims), our results show that $\mathcal{P}_M$ changes less than $0.1\%$ when $L$ reaches the length of $10^4$. This might suggest that there is no advantage in using the simplistic self-replication mechanism beyond $L\sim 10^4$, in reasonable agreement with the transition from RNA-based to DNA-based life.

{\bf Acknowledgments:} Partial support from Brazilian agencies PROPP/UFOP, CNPq, FAPEMIG and FAPES is acknowledged.

%\bibliography{Refs}


\end{document}